# Modulational-instability-free pulse compression in anti-resonant hollow-core photonic crystal fiber

FELIX KÖTTIG,[1,*] FRANCESCO TANI[1], AND PHILIP ST.J. RUSSELL[1,2]

[1]Max Planck Institute for the Science of Light and [2]Department of Physics, Friedrich-Alexander-Universität, Staudtstr. 2, 91058 Erlangen, Germany
*Corresponding author: felix.koettig@mpl.mpg.de

**Gas-filled hollow-core photonic crystal fiber (PCF) is used for efficient nonlinear temporal compression of femtosecond laser pulses, two main schemes being direct soliton-effect self-compression, and spectral broadening followed by phase compensation. To obtain stable compressed pulses, it is crucial to avoid decoherence through modulational instability (MI) during spectral broadening. Here we show that changes in dispersion due to spectral anti-crossings between the fundamental core mode and core wall resonances in anti-resonant-guiding hollow-core PCF can strongly alter the MI gain spectrum, enabling MI-free pulse compression for optimized fiber designs. In addition, higher-order dispersion can introduce MI even when the pump pulses lie in the normal dispersion region.**

Modulational instability (MI), the parametric amplification of weak perturbations in nonlinear systems, is a phenomenon that can be observed in many different fields, for example, ocean waves, biological systems, chemical reactions, plasma physics and nonlinear optics [1]. One of the most prominent MI mechanisms is the Benjamin-Feir instability [2], which causes spontaneous exponential growth of spectral sidebands and periodic modulation of the homogeneous steady-state solution. In the context of nonlinear fiber optics, Benjamin-Feir instabilities arise from the interplay between Kerr nonlinearity and (typically anomalous) dispersion [3], initiating, for example, supercontinuum generation in photonic crystal fiber (PCF) [4,5]. Seeded by quantum noise, MI generates an incoherent train of pulses. Avoiding the onset of MI is therefore critical when designing fiber-based pulse compression systems. In many cases, this requires pumping in the normal dispersion region (positive group velocity dispersion [GVD]). However, this is not always practical, and scalar MI can still occur for example through higher-order dispersion [6], making some kind of mechanism desirable for effective suppression of MI.

Here we propose dispersion-engineering of anti-resonant-guiding hollow-core PCF for MI-free pulse compression. For nonlinear temporal compression of intense femtosecond laser pulses, spectral broadening via self-phase modulation (SPM) in gas-filled anti-resonant kagomé-type or single-ring PCF is particularly attractive as it is highly efficient and scalable to megahertz repetition rates [7–10]. In these fibers, broadband low-loss guidance bands (loss <10 dB km$^{-1}$ [11,12]) are interspersed by spectral anti-crossings between the fundamental core mode and core wall resonances, which strongly alter the dispersion and introduce high loss. The anti-crossings can be modelled using the approach introduced by Archambault et al. [13] and their frequency can be estimated as $f_m = mc/(2t\sqrt{n_{\text{wall}}^2 - n_{\text{gas}}^2})$, where $t$ equals the capillary wall thickness in single-ring PCF and the core wall thickness in kagomé-type PCF, $n_{\text{wall}}$ and $n_{\text{gas}}$ are the refractive indices of the core wall and the filling gas, $m$ is the order of the wall resonance, and $c$ is the speed of light in vacuum. Although spectral broadening should occur predominantly within the low-loss guidance bands, the change in dispersion due to the anti-crossings can still sensitively influence pulse compression and supercontinuum generation as well as parametric processes such as dispersive wave emission and four-wave mixing [14–16]. In addition, it has also been shown that localized loss bands can induce MI [17]. For pulse compression, avoiding MI is of the utmost importance, since it degrades the coherence of the compressed pulses.

Here we introduce a simple analytical model that accurately predicts the MI gain spectrum in anti-resonant PCF, including the influence of the anti-crossings. Based on this model, we identify values of wall thickness that shift the pump pulses into the positive GVD region, thus effectively suppressing the onset of MI. The predicted MI-free spectral broadening is confirmed by numerical simulations of pulse propagation. Finally, we demonstrate that higher-order dispersion can induce MI even when technically pumping in the positive GVD region (which is normally assumed to be MI-free)—an effect that has previously been reported in solid-core PCF [6].

In this paper we analyze a typical setup for nonlinear temporal compression of pulses from an ytterbium laser (1 μm central wavelength, 350 fs pulse duration, 15 μJ pulse energy) via SPM-based spectral broadening in gas-filled PCF and subsequent phase compensation (for example using chirped mirrors) [8,18]. The fiber is a 1-m-long single-ring PCF with 50 µm core diameter filled with 10 bar argon [Fig. 1(a)] (similar results apply to kagomé-type PCF [19]). The fiber cladding consists of seven thin-walled capillaries with 25 µm diameter and variable wall thickness (the wall thickness is varied for dispersion-tuning as shown later). To avoid time-consuming finite element modelling (FEM) of the fiber modes, we approximate the real part of their effective refractive index (modal index) via the model from Zeisberger and Schmidt [20] and their loss via the model from Vincetti and Rosa [21]. This gives reasonable agreement with FEM data in the relevant spectral region from 200 to 400 THz [Fig. 1(b),(c)].

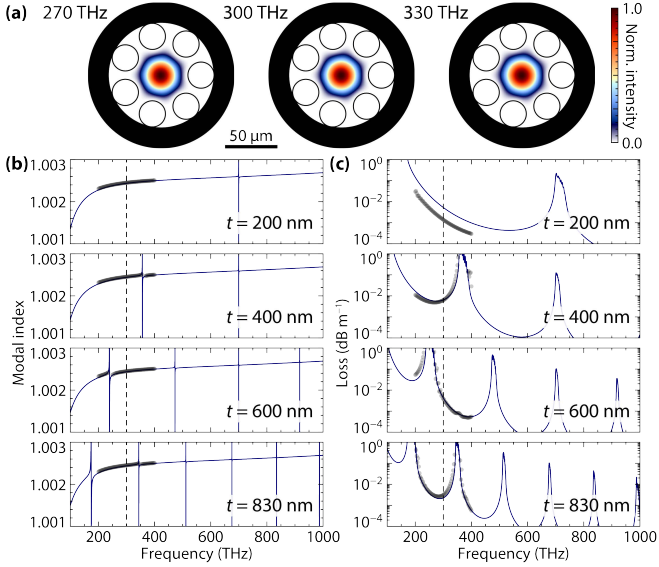

**Fig. 1.** (a) Structure of the single-ring PCF (50 µm core diameter, 400 nm capillary wall thickness, filled with 10 bar argon) and intensity distribution of the fundamental core mode at frequencies of 270, 300 and 330 THz. (b) Modal index [20] and (c) loss [21] of the fiber, for capillary wall thicknesses $t$ = 200, 400, 600 and 830 nm. The gray dots are data from finite element modelling. The dashed lines denote the pump frequency used in this work (300 THz).

In the following, we present a complete model for the MI gain spectrum in anti-resonant PCF, which we then use to analyze the influence of the fiber anti-crossings. To this end, we consider the generalized nonlinear Schrödinger equation in a co-moving reference frame, including full dispersion and Kerr nonlinearity (neglecting the shock effect and Raman nonlinearity):

$$\partial_z U(z,t) = DU(z,t) + i\gamma |U(z,t)|^2 U(z,t), \quad (1)$$

where $U$ is the complex pulse envelope, scaled such that $|U|^2$ is the instantaneous power in units of Watts, $z$ is the propagation distance, $t$ is time, $D$ is the dispersion operator in time domain and $\gamma$ is the nonlinear fiber parameter [3]. The spatial profile of the $LP_{01}$-like fundamental mode was approximated by a $J_0$ Bessel function and the dispersion and nonlinearity of the gas were taken from [22] and [23]. The steady-state solution of Eq. (1) is given by $U(z,t) = \sqrt{P_0} \exp(i\gamma P_0 z)$, where $P_0$ is (peak) power. Following the standard linear stability analysis (see, for example, [3]), we add a perturbation $u(z,t)$ to the steady-state solution of Eq. (1)

$U(z,t) = \left(\sqrt{P_0} + u(z,t)\right)\exp(i\gamma P_0 z)$. Linearizing in $u$ and using the ansatz $\hat{u}(z,\omega) = \hat{u}_1 \exp(K_1 z) + \hat{u}_2 \exp(K_2 z)$, where $\omega$ is the angular frequency and a hat denotes frequency domain representation, the MI wavevector turns out to be (an equivalent form can be found in [24])

$$K_{1,2} = (\hat{D} + \hat{D}^*)/2 \pm \sqrt{(\hat{D} - \hat{D}^*)^2/4 + i\gamma P_0(\hat{D} - \hat{D}^*)}. \quad (2)$$

The spectrum of the MI power gain is then given by $G = 2\Re(K)$, where $\Re$ denotes the real part. In the co-moving reference frame of Eq. (1), the dispersion operator in the frequency domain is given by $\hat{D} = i(\beta(\omega) - \beta(\omega_0) - \beta_1(\omega_0)(\omega - \omega_0))$, where $\beta = \omega n_{\text{eff}}/c$ is the modal propagation constant, $n_{\text{eff}}$ the real-valued modal index, $\omega_0$ the central frequency of the pump and $\beta_1 = \partial_\omega \beta$. $\hat{D}^*$ can be calculated as $\hat{D}^* = \mathcal{F}\left([\mathcal{F}^{-1}(\hat{D})]^*\right) = \left(\hat{D}(-\omega)\right)^*$, where $\mathcal{F}$ denotes the Fourier transform. In the model the complete fiber dispersion is used, without Taylor expansion, allowing exact modelling of complex dispersion curves, including the fiber anti-crossings.

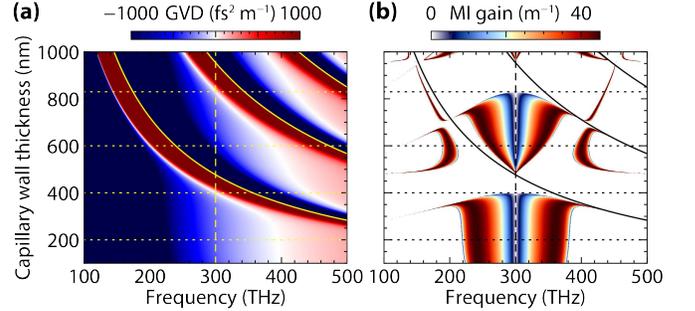

**Fig. 2.** (a) Group velocity dispersion (GVD) and (b) modulational instability (MI) gain for a fiber with 50 µm core diameter filled with 10 bar argon and pumped by 350 fs, 15 µJ Gaussian pulses centered at 300 THz (vertical dashed lines). The solid lines mark the locations of the anti-crossings. The horizontal dotted lines denote capillary wall thicknesses of $t$ = 200, 400, 600 and 830 nm as used in the simulations in Fig. 3. At $t$ = 200 nm, the MI gain peaks at 241 and 359 THz and at $t$ = 600 nm, it peaks at 199, 263, 337 and 401 THz.

Figure 2 shows the GVD $\beta_2 = \partial_\omega^2 \beta$ of the fiber and the MI gain spectrum for a pump at 300 THz (1 µm wavelength) as function of capillary wall thickness. The fiber anti-crossings add very strong positive (negative) GVD on their low-frequency (high-frequency) side, which locally changes the otherwise hollow capillary-like dispersion. For very small capillary wall thickness ($t$ < 200 nm), the anti-crossing frequencies are far away from the pump and the dispersion is close to that of a hollow capillary over a very wide frequency range, with negative GVD at the pump frequency. In this case, a classical MI gain spectrum is seen, with lobes symmetrically placed about the pump frequency. For larger $t$, on the other hand, the anti-crossings shift towards the pump frequency, where they strongly modify the GVD landscape, creating regions of negative and positive GVD. This eliminates the MI gain lobes seen at smaller values of $t$, while creating new ones at the same time. Interestingly, there are extended regions of zero MI gain for certain values of $t$, for example 400 nm. These regions arise mainly through strong dispersion introduced by the anti-crossings, which leads to positive

GVD at the pump frequency (note that positive GVD is not always sufficient to suppress MI, as we will discuss later). Although this MI suppression is caused by the strong change in dispersion in the vicinity of the fiber anti-crossings, the usable bandwidth over which the spatial modes are undisturbed and the loss is tolerable is still >60 THz (Fig. 1), which is more than enough for pulse compression to sub-30 fs duration. The predicted regions of zero MI gain are potentially interesting for pulse compression, which will be analyzed in the following.

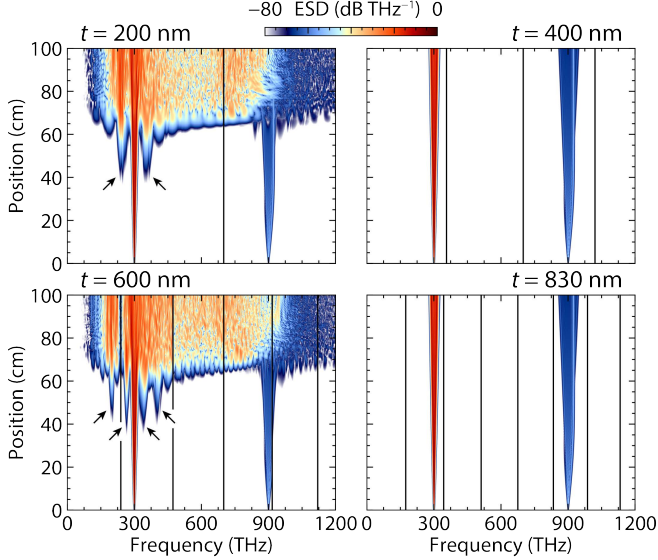

**Fig. 3.** Simulated pulse propagation in a fiber with 50 μm core diameter filled with 10 bar argon and pumped by 350 fs, 15 μJ Gaussian pulses centered at 300 THz, for capillary wall thicknesses $t$ = 200, 400, 600 and 830 nm. The solid lines denote the spectral locations of the fiber anti-crossings. For $t$ = 200 nm, modulational instability sidebands (indicated by the arrows) appear at ~240 and ~360 THz and for $t$ = 600 nm, they appear at ~200, ~265, ~340 and ~400 THz.

To first order, spectral broadening due to SPM scales with the product $\gamma P_0 L$, where $L$ is the fiber length. For stable operation at high repetition rates and average power, it is desirable to reduce $\gamma$ (i.e., the gas pressure) and increase $L$ [10]. However, this is limited by the onset of MI when pumping in the negative GVD region, leading to pulse breakup in long fibers. Optimizing the fiber design, as proposed here, provides a promising route to pulse compression at high powers in anti-resonant PCFs, since it allows use of lower gas pressures and longer propagation lengths, while effectively avoiding MI. For the different MI regimes delineated in Fig. 2, we rigorously modeled pulse propagation using a carrier-resolved unidirectional pulse propagation equation, including fiber loss and the shock effect [25]. Quantum noise was included as one photon per frequency mode (spectral bin of width 0.2 THz) with random phase [4] (for better comparability, the same noise was used in all simulations). Figure 3 shows the evolution of the pulse spectrum along the fiber for different values of $t$. As predicted from the MI gain spectrum in Fig. 2, for $t$ = 200 and 600 nm MI sidebands appear at $z \sim$ 40 cm, with spectral positions that agree very well with the model. Upon further propagation, they quickly lead to pulse breakup and extremely broadband, incoherent, supercontinuum generation, which must be avoided for pulse compression. At $t$ = 400 and 830 nm, on the other hand, no MI sidebands are visible over an 80 dB dynamic range, and clean SPM-based spectral broadening is observed. In this case, the pulse can be compressed from the initial 350 fs to ~24 fs by introducing a group delay dispersion of −1380 fs$^2$, with less than 0.15% propagation loss. Since the suppression of MI is caused by changes in dispersion, switching off fiber loss in the simulations does not significantly change the dynamics. Because of the strong dispersion introduced by the anti-crossings, the method can be applied also to fibers with different core diameters. We verified this for pulse compression at lower energy (5 μJ) in a fiber with core diameter 30 μm filled with 10 bar of argon, as well as at higher energy (45 μJ) in a fiber with 70 μm core filled with 5 bar of argon.

Another interesting effect can be observed in Fig. 3. While an anti-crossing on the high-frequency side of the pump ($t$ = 400 nm) can suppress MI, an anti-crossing on the low-frequency side ($t$ = 600 nm) leads to supercontinuum generation, similar to what is seen when the anti-crossings are far away from the pump ($t$ = 200 nm). Apparently, MI-based supercontinuum generation does not require very small capillary wall thickness in this case. Indeed, thicker walls even seem to improve the extent and flatness of the supercontinuum, as previously reported [14]. This is in stark contrast to coherent soliton-based pulse compression and ultraviolet light generation, where the anti-crossings must be sufficiently far from the pump frequency for the process to be efficient [16], which usually does require relatively thin-walled capillaries that can be difficult to fabricate.

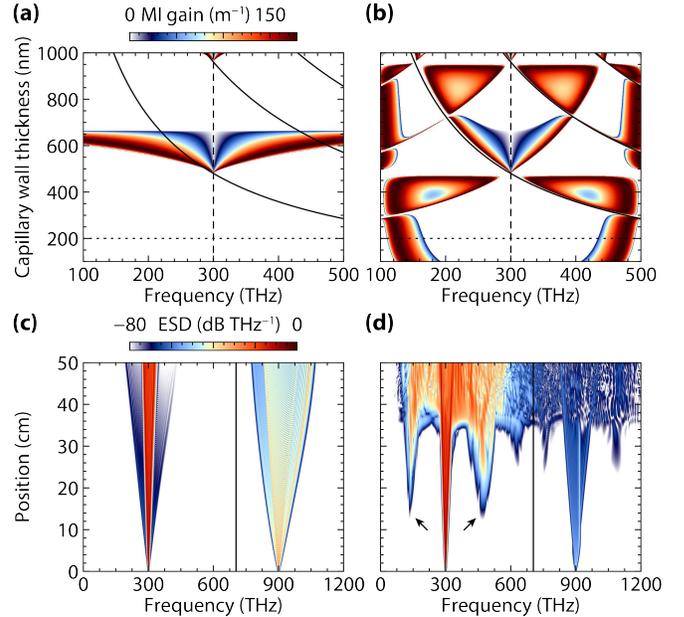

**Fig. 4.** Fiber with 50 μm core diameter filled with 40 bar argon and pumped by 350 fs, 15 μJ Gaussian pulses centered at 300 THz (vertical dashed lines). **(a),(b)** Modulational instability (MI) gain and **(c),(d)** simulated evolution of the pulse spectrum, **(a,c)** considering only second-order dispersion and **(b,d)** including the full dispersion. The solid curves mark the spectral locations of the fiber anti-crossings. The horizontal dotted lines correspond to $t$ = 200 nm as used in **(c),(d)**. In **(c)**, the weak broadband background in the pump spectrum is due to cross-phase modulation with the strong third and fifth harmonics. In **(d)**, MI sidebands (indicated by the arrows) appear at ~135 and ~475 THz, while the model in **(b)** predicts 126 and 474 THz.

When higher-order dispersion is neglected in the simulations, scalar MI can only develop when pumping in the negative GVD region. When pumping in the positive GVD region, however, higher-order dispersion can lead to MI [6]. For anti-resonant PCF-based pulse compression systems, this means that the naïve assumption that MI can be suppressed by pumping in the positive GVD region must be carefully checked. Figure 4 compares the MI gain spectrum and the pulse evolution when pumping in the positive GVD region (40 bar argon pressure). Using a Taylor expansion of the dispersion operator up to second order results in the expected suppression of MI at small capillary wall thickness, when the fiber anti-crossings have only a small effect on the dispersion in the vicinity of the pump frequency [Fig. 4(a),(c)]. Including the full dispersion, on the other hand, recovers MI gain bands that are spaced far from the pump [Fig. 4(b),(d)], similar to what has been observed in solid-core PCF when pumping close to the zero-dispersion frequency [6]. Furthermore, the MI gain spectrum does not predict any useful regions of zero MI gain by tuning the capillary wall thickness in this case. Consequently, also in anti-resonant PCF, simply pumping in the positive GVD region might not be sufficient to avoid MI, and a careful parameter analysis is necessary for the design of pulse compression systems.

In conclusion, spectral anti-crossings with cladding wall resonances have a strong effect on the dispersion in anti-resonant PCF, with important implications for SPM-based pulse compression. An analytical model and numerical simulations predict that optimizing the fiber parameters (in particular the wall thickness in the anti-resonant elements) can offer MI-free pulse compression at low gas pressures, which is advantageous for stable operation at high average power. The approach can also be used to optimize other phase-matched processes, for example to frustrate coherent Raman gain suppression [26] by designing an anti-crossing to dephase the anti-Stokes band. The results are applicable at any pump laser wavelength where the fiber transmits, for example 800 nm (Ti-sapphire lasers). Finally, in anti-resonant PCF, even when technically pumping in the positive GVD region, MI can be induced by higher-order dispersion, preventing coherent spectral broadening and stable pulse compression.